\documentstyle [aps]{revtex}
\begin{document}
\draft
\title{
R\"ontgen interaction term makes dipole approximation more divergent
}
\author{Marek Czachor~\cite{*}}
\address{
Research Laboratory of Electronics\\
 Massachusetts Institute of Technology,
 Cambridge, MA 02139}
\author{Kazimierz Rz\c a\.zewski}
\address{
Centrum Fizyki Teoretycznej PAN, \\ Al. Lotnik\'{o}w 32/46,
02-668 Warszawa, Poland
}
\maketitle
\begin{abstract}
It is shown that inclusion of the R\"ontgen term in the
interaction Hamiltonian
leads to an additional divergency, which does not occur in the
standard approach to the dipole approximation.
The physical reason of this divergency is the too rapid growth
with frequency of
the ``coupling constants" resulting from the R\"ontgen
interaction. Therefore even in calculation of the
radiation pattern one has to explicitly introduce  a formfactor
if the atomic mass $M<\infty$.
\end{abstract}

\pacs{42.50.Vk, 42.50.Wm,32.70.Fw, 32.80.Pj}

The usage of the conventional dipole interaction $-{\vec
d}\cdot{\vec E}({\vec r})$ for moving atoms has been recently
criticised by Wilkens \cite{W1,W2}, who argued that one has
to include the R\"ontgen interaction term if one wants to
describe the spontaneous emission of a moving atom trustworthy
to first order in atomic velocity. Since the basic feature of
the
R\"ontgen term is the momentum dependence of the atom---field
``coupling constants" it seems that its
inclusion
could modify a
dependence of the spectrum of emitted radiation on the shape of
the center of mass atomic wave packet.

The fact that such a dependence is present follows both from
the calculations based on the standard $-{\vec
d}\cdot{\vec E}({\vec r})$ term \cite{R} and the experiment
of
Robert {\em et al.\/}\cite{Baudon}. However, the dependence
observed experimentally is much stronger than the one found in
theoretical calculations.

The aim of this Brief Report is to investigate the role of the
R\"ontgen interaction for spectral properties of light
emitted spontaneously by an atomic wavepacket.
We will see that the obtained modification of the spectrum
does not significantly differ from the results obtained in the
standard way provided one uses the infinite atomic mass limit.
Otherwise the probability of spontaneous emission in a given
direction is represented by a divergent integral.

We consider the Hamiltonian
\begin{equation}
H=\frac{\hat {\vec p}\,{^2}}{2M}+H_A+H_F+H_{AF}
\end{equation}
where the first term is the atomic center-of-mass kinetic
energy,
\begin{eqnarray}
H_A&=&\frac{1}{2}\hbar\omega_0\sigma_3,\\
H_F&=&\sum_{{ k},\lambda}\hbar\omega_{ k}a^{\dag}_{{
k}\lambda} a_{{ k}\lambda},
\end{eqnarray}
and the interaction Hamiltonian equals \cite{W2}
\begin{equation}
H_{AF}= -i\sum_{{ k},\lambda} g_{{ k}\lambda}(\hat {\vec p})
e^{i{\vec k}\cdot\hat {\vec r}}\sigma_+a_{{ k}\lambda}+{\rm
h.c.}
\end{equation}
The operator valued coupling is
\begin{equation}
g_{{ k}\lambda}(\hat {\vec p})={\cal E}_{ k}d\Bigl\{
({\vec e}_d\cdot {\vec e}_{{ k}\lambda})
\Bigl[1-{\vec n}_{ k}\cdot\hat {\vec \beta} +
\hbar\omega_{{ k}}/(2Mc^2)\Bigr]
+\bigl({\vec e}_d\cdot{\vec n}_{ k}\bigr)\bigl({\vec e}_{{
k}\lambda}
\cdot\hat {\vec \beta}\bigr)\bigr)
\Bigr\},\label{CC}
\end{equation}
with $\hat {\vec{\beta}}=\hat {\vec p}/(Mc)$, ${\vec e}_d$ the
unit vector in the direction of the atomic
dipole moment, ${\vec e}_{{ k}\lambda}$, $\lambda=1,2$ the
unit vectors in the
polarization directions, $d$ the value of the dipole moment, and
${\cal E}_{ k}=\sqrt{\hbar\omega_{ k}/2\epsilon_0{\cal
V}}$  the electric field strength per photon in the
quantization volume $\cal V$. It should be stressed that now the
``coupling constants" are given by {\em time dependent\/}
operators (the recoil makes the atomic momentum time dependent).

Following the standard Weisskopf-Wigner procedure \cite{R} we
assume that the state of the atom--field system is
\begin{equation}
|\psi\rangle = \int d^3p\,\alpha({\vec p})|{\vec p},+,0\rangle+
\sum_{{ k},\lambda}
\int d^3p\,\beta({\vec p},{\vec k},\lambda)|{\vec p}-\hbar{\vec
k},-,{\vec k},\lambda\rangle \label{st}
\end{equation}
leading to the Schr\"odinger equation
\begin{eqnarray}
i\hbar\dot \alpha({\vec p})&=&\Biggl(\frac{{\vec p}^2}{2M} +
\frac{1}{2}\hbar\omega_0\Biggr)\alpha({\vec p})
-i\sum_{{ k},\lambda}
g_{{ k}\lambda}({\vec p})\beta({\vec p},{\vec k},\lambda)\\
i\hbar\dot \beta({\vec p},{\vec k},\lambda)&=&
\Biggl(\frac{({\vec p}-{\vec k})^2}{2M} -
\frac{1}{2}\hbar\omega_0+\hbar\omega_{ k}\Biggr)
\beta({\vec p},{\vec k},\lambda)
+i
g_{{ k}\lambda}({\vec p}+\hbar{\vec k})\alpha({\vec p}).
\end{eqnarray}

 The solutions of the
Schr\"odinger equation (in  the single pole approximation) are
\begin{eqnarray}
\alpha_t({\vec p})&=&\alpha_0({\vec p})e^{-z_0t},\\
\beta_t({\vec p},{\vec
k},\lambda)&=&-\frac{1}{\hbar}\alpha_0({\vec
p})
g_{{ k}\lambda}({\vec p}+\hbar{\vec k})\frac{e^{-z_0t}
-e^{-z_{ k}t}}{z_0-z_{ k}},\label{10}
\end{eqnarray}
where
\begin{eqnarray}
z_0&=& \frac{i}{\hbar}\Bigl(\frac{{\vec
p}^2}{2M}+\frac{1}{2}\hbar\omega_0\Bigr) +\frac{\gamma_0}{2}
\label{11}\\
z_{ k}&=& \frac{i}{\hbar}\Bigl(\frac{({\vec
p}-\hbar{\vec k})^2}{2M}-\frac{1}{2}\hbar\omega_0
+\hbar\omega_{ k}\Bigr)\label{12}
\end{eqnarray}
with the Lamb shift included in the level spacing.
The rate of spontaneous emission $\gamma_0$, in the infinite
mass limit (and only then!), is the same as in ordinary theory.

Consider now, for simplicity, the probability of emitting a photon
whose wave-vector is perpendicular to the atomic dipole moment.
For $t\gg 1/\gamma_0$ we find
\begin{equation}
P_{ k}=\frac{d^2}{2\epsilon_0\hbar {\cal V}}\int d^3p\,
|\alpha_0(\vec p)|^2 \frac{\omega_{ k}
\bigl(1-\vec n_{ k}\cdot\vec
\beta-
\hbar\omega_{ k}/(2Mc^2)\bigr)^2}
{\bigl(\omega_0-\omega_{ k}(1-\vec n_{ k}\cdot\vec
\beta)-\hbar\omega_{ k}^2/(2Mc^2)\bigr)^2+\gamma_0^2/4}.\label{13}
\end{equation}
It is  interesting that the same dependence on the atomic wave packet
would be found if instead of the atomic pure state we considered
a mixed state density matrix corresponding to the same initial probability
distribution of atomic momenta. This is one way of seeing that
the modification of the radiation patern does not result from
an interference of light but rather from the distribution of
the Doppler shifts in the initial wavepacket.

The recoil term in the numerator in (\ref{13}) (as opposed to
the one in the
denominator) is a direct consequence
of the R\"ontgen interaction. It appears  because of two reasons.
First,
the identical term is present already
 in the coupling (\ref{CC}) and represents the recoil contribution to
the coupling. However, even if we eliminated this
term from the very beginning by assuming the infinite mass of the atom,
the term would reappear through the Doppler part of (\ref{CC})
since,
additionally, $\beta_t({\vec p},{\vec k},\lambda)$ is
proportional to $g_{{ k}\lambda}({\vec p}+\hbar{\vec
k})$ and the momentum shift $\vec p\to \vec p +\hbar\vec k$
would lead to an analogous contribution but twice bigger and with opposite
sign. All these additional expressions do not occur in calculations based
on the ordinary dipole interaction, where we do not find  the momentum
shift in the coupling (the Doppler contribution), and the recoil modification
is simply absent.
Now, if we consider the probability of
 emission in a given
direction,
\[
\frac{{\cal V}}{(2\pi c)^3}\int_0^\infty d\omega_{
k}\,\omega_{ k}^2P_{ k},
\]
we find that the integral is divergent. The divergence
follows from the obvious fact that now the rank of the polynomial
in the numerator is greater than in the denominator.
To eliminate this pathology one must
put $\hbar\omega_{ k}^2/(2Mc^2)=0$ in (\ref{CC}) and $g_{{
k}\lambda}({\vec p}+\hbar{\vec k})$, which can be obtained in a
mathematically consistent way only by elimination of the first additional
two terms in (\ref{CC}), or by a formfactor, but then both the
shape of the
emission line and the atomic lifetime would explicitly depend on its choice.
The last part of the R\"ontgen term will not cause problems here, as
the momentum shift is in the direction perpendicular to the polarization
vectors.

It is interesting to compare (\ref{13}), which follows from the
Schr\"odinger equation, with the calculations based on the Fermi golden
rule chosen in \cite{W2} in the form
\begin{equation}
d\Gamma_p= \frac{2\pi}{\hbar}\sum_{k,\lambda}|g_{k\lambda}(\vec p)|^2
\delta\Bigl\{\hbar\omega_k\Bigl[1-\frac{\vec n_k\cdot\vec p}
{Mc}+\frac{\hbar\omega_k}{2Mc^2}\Bigr]   -\hbar\omega_0\Bigr\}
\delta(\vec n - \vec n_k)d^2\vec n.\label{F}
\end{equation}
Now the limit $M\to\infty$ in (\ref{F}) and (\ref{CC}) is essentially
equivalent to neglecting the recoil terms in (\ref{13}).
But the rule that follows from the solution (\ref{10}) would be rather
\begin{equation}
d\Gamma_p= \frac{2\pi}{\hbar}\sum_{k,\lambda}|g_{k\lambda}(\vec p+
\hbar\vec k)|^2
\delta\Bigl\{\hbar\omega_k\Bigl[1-\frac{\vec n_k\cdot\vec p}
{Mc}+\frac{\hbar\omega_k}{2Mc^2}\Bigr]   -\hbar\omega_0\Bigr\}
\delta(\vec n - \vec n_k)d^2\vec n.\label{F'}
\end{equation}
We can now put $M\to\infty$, but this must be done {\em before evaluation
of the sum over momenta\/}. Otherwise the expression will lead to the
discussed  divergency. (\ref{F'}) shows again that putting $1/M=0$ in
(\ref{CC}) will not eliminate the difficulty.

Summarizing, the R\"ontgen term can be applied to infinitely heavy atoms,
provided one knows at which stage of the calculations to neglect the
terms involving $1/M$.

\bigskip
The authors would like to thank Martin Wilkens for his comments.

\end{document}